\documentclass[twocolumn]{jpsj2}
\usepackage{txfonts}
\usepackage{graphicx}

\title{Unidirectional Electronic Structure in the Parent State of Iron-Chalcogenide Superconductor Fe$_{1+\delta}$Te}

\author{\textsc{Tadashi Machida}$^{1,2}$\thanks{E-mail address:tmachida@rs.tus.ac.jp},
 \textsc{Kazuho Kogure}$^{1}$, \textsc{Takuya Kato}$^{1}$, \textsc{Hiroshi Nakamura}$^{1}$, \textsc{Hiroyuki Takeya}$^{3}$, \textsc{Takashi Mochiku}$^{2}$, \textsc{Shuuichi Ooi}$^{2}$, \textsc{Yoshikazu Mizuguchi}$^{4}$, \textsc{Yoshihiko Takano}$^{3}$, \textsc{Kazuto Hirata}$^{2}$ and \textsc{Hideaki Sakata}$^{1}$}

\inst{$^{1}$Department of Physics, Tokyo University of Science, Tokyo 162-8601, Japan
\\$^{2}$Superconducting Properties Unit, National Institute for Materials Science, 1-2-1 Sengen,Tsukuba, Ibaraki 305-0047, Japan
\\$^{3}$Superconducting Wires Unit, National Institute for Materials Science, 1-2-1 Sengen,Tsukuba, Ibaraki 305-0047, Japan
\\$^{4}$Electrical and Electronic Engineering, Tokyo Metropolitan university, 1-1 Minami-osawa, Hachioji, Tokyo 192-0397, Japan
}
\abst{We use scanning tunnelling microscopy and spectroscopy to explore the electronic structure of Fe$_{1.07}$Te which is the parent compound of the iron-chalcogenide superconductors.
A unidirectional electronic structure with a period of $a_{0}$ (where $a_{0}$ is the lattice constant) along the $a$-axis is observed.
The unidirectional pattern is visible within an energy range from approximately -200 to +130 meV.
Since the direction of the unidirectional electronic structure coincides with those of the underlying antiferromagnetic and the predicted orbital orders, it is presumable that the observed unidirectional structure is closely tied to these orders and is peculiar to the parent state in iron-based superconductors.}

\kword{Scanning tunneling microscope and spectroscopy, Iron-based superconductors, Iron-chalcogenide}

\begin{document}
\maketitle

\section{Introduction}
Strong electronic correlation yields several orders, which attract much attention in solid state physics since the complex entanglement of these orders creates many interesting and unique phenomena.
One of the most intriguing examples of such phenomena is the high temperature superconductivity in the cuprates with antiferromagnetic (AFM) order and several types of electronic orders. \cite{Zaanen,Kivelson,Sachdev,MKato,Hanaguri_1,Kohsaka,Kohsaka_1,Lawler,Parker,TMachida} Another example is the colossal magnetoresistance in the manganites with magnetic, orbital, and charge order \cite{Perring,Zimmermann,Kawano,Staub,Murakami}.
The recently discovered iron-based superconductors (SCs) \cite{Kamihara,Rotter_1,Rotter_2,Kasahara,Tapp,XCWang,XZhu,Mizuguchi_1,Hsu} also have AFM \cite{Cruz,Huang,WBao,SLi} and orbital orders \cite{Jensen,Shimojima,Kruger,CCChen,CCLee,Lv,Turner} in their parent compounds.
The chemical doping of the parent compounds, inducing the fluctuation of these orders, causes the high temperature superconductivity in iron-based SCs: it has been theoretically argued that the electron pairing is mediated by the spin fluctuation \cite{Mazin,Kuroki,FWang,Cvetkovic} or the orbital fluctuation \cite{JZhang,TSaito,Kontani}.
Thus, in iron-based materials, the elucidation of the effect of these orders on the electronic structure is necessary to understand the pairing mechanism.

Recently, scanning tunnelling spectroscopy (STS) experiments based on scanning tunneling microscope (STM) have revealed unidirectional (C$_2$ symmetric or nematic) electronic structures in the parent compounds CaFe$_{1.94}$Co$_{0.06}$As$_{2}$ (122 system)\cite{TMChuang} and LaOFeAs (1111 system)\cite{XZhou}.
Even though the physical picture regarding such unidirectional structures is still under debate \cite{Mazin_1,HZhai,WCLee}, the unidirectional electronic structure appears to be strongly tied to the underlying AFM and/or orbital orders, because the direction of the unidirectional features coincides with those of these orders.
To verify whether the electronic unidirectionality is really related to the underlying orders, it is quite important to investigate the electronic structures in the parent materials containing AFM and orbital orders with different spin and orbital configurations.

\begin{figure}[b]
\begin{center}
\includegraphics[width=7cm]{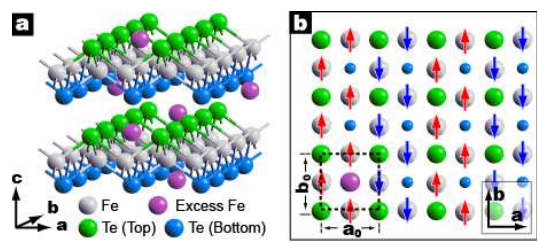}
\end{center}
\caption{
(Color online) (a) and (b) Schematic illustrations of the crystal structure of Fe$_{1+\delta}$Te.
Red and blue arrows indicate the spin configurations at Fe sites in FeTe layer. The direction of AFM order coincides with $a$-axis. }
\label{Fig_Schem}
\end{figure}

One of the candidates containing a different configuration is Fe$_{1+\delta}$Te, which is the parent compound of the iron-chalcogenide SCs.
This material exhibits the structural transition from the tetragonal to the monoclinic structure at the structural transition temperature ($T_{\mathrm{s}}$), being accompanied by the bicollinear AFM order.
In this AFM state, the spins are aligned antiferromagnetically along a diagonal direction (the crystal $a$-axis) and ferromagnetically along the other diagonal direction on the Fe rectangular lattice (the $b$-axis), as illustrated in Fig. \ref{Fig_Schem}(b) \cite{WBao,SLi}.
According to recent theoretical work on Fe$_{1+\delta}$Te, an orbital order lies in this material, and its direction is different from those in the 122 and 1111 systems.

In this paper, we demonstrate STS studies in Fe$_{1+\delta}$Te. This material has the simplest crystal structure, as shown in Fig. \ref{Fig_Schem}, and is most suitable for STS experiments in all of iron-based SCs because of the ease of preparing a clean surface by cleaving a sample, the electric neutrality of the cleaved surface, and the absence of surface reconstruction (that is, the reproducible appearance of the chalcogen layer) \cite{Hanaguri_2,Kato,Massee}.
We demonstrated the existence of the unidirectional electronic structure along the crystal $a$-axis or spin antiparallel direction in the bicollinear AFM order [Fig. \ref{Fig_Schem} (b)] \cite{WBao} with a period of $a_{0}$ (where $a_{0}$ is the lattice constant the along $a$-axis).
The unidirectional pattern is visible within an energy range from about -200 to +130 meV.
Even though the direction of the underlying AFM and orbital orders in this material is different from those in other systems, the direction of the observed electronic unidirectionality coincides with those of the underlying orders, as in the 122 and 1111 systems.
Therefore, the electronic unidirectionality seems to be related to the underlying orders and to be one of the inherent features in the parent materials of iron-based SCs.

\begin{figure}[tb]
\begin{center}
\includegraphics[width=7cm]{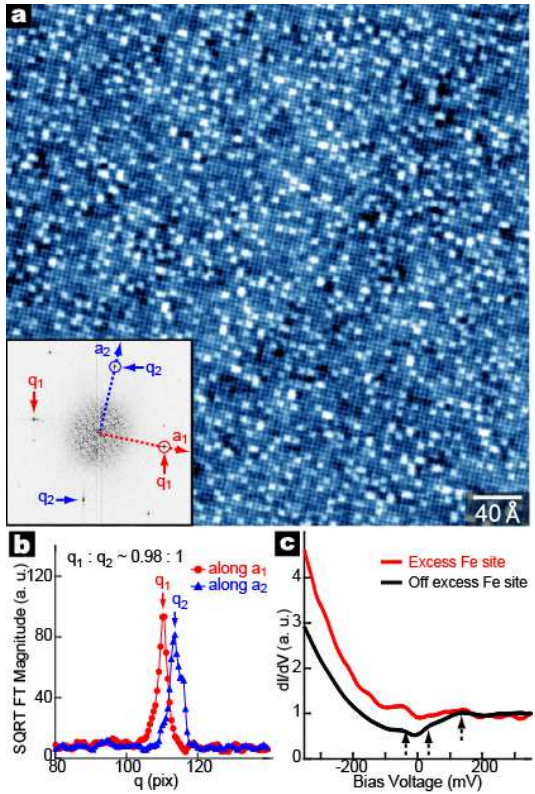}
\end{center}
\caption{(Color online)
(Color online) (a) An STM image taken at $V_{\mathrm{B}}$ = -500 mV and $I_{\mathrm{set}}$ = 500 pA on a 433-\AA\ square field of view (FOV).
Inset shows Fourier transform image of (a). Four solid arrows (two red and two blue arrows) represent the spots corresponding to the atomic lattice of the exposed top Te atoms.
(b) Line cuts of the FT image along $a_{1}$ (red circles) and $a_{2}$ (blue triangles) directions. The peak positions in these line cuts are indicated by red and blue arrows and named as $q_{1}$ and $q_{2}$, respectively.
(c) Tunneling spectrum obtained by averaging the spectra on several excess Fe sites (red) and the spatially averaged spectrum all region except excess Fe sites (black).
}
\label{Fig_Topo}
\end{figure}

\section{Experiment}
The samples used in this study were cut from a single crystal of Fe$_{1+\delta}$Te grown by a melting-growth technique \cite{XtalGrowth}. The molar ratio of the grown crystal was determined to be Fe$_{1.07}$Te by inductively coupled plasma mass spectroscopy.
The AFM and structural transition temperatures $T_{\mathrm{s}}$ were determined to be 64 K by SQUID magnetization and resistivity measurements.
The STS measurements were performed in a helium gas environment at 4.2 K on an atomically flat surface, as shown in Fig. \ref{Fig_Topo}(a), prepared by cleaving the samples \textit{in-situ} at 4.2 K.
At this temperature, the crystal structure of the sample is monoclinic: the lattice constants along the $a$- and $b$-axes are not equivalent.
We measured the $I$-$V$ characteristics and obtained the differential conductance $G(\boldsymbol{r}, E = eV_{\mathrm{B}}) = dI/dV$ by numerical differentiation.

\section{Results}
\subsection{Typical surface structure and identification of crystal $a$- and $b$-axes}
In a typical STM image of the sample surface, a square lattice structure is observed, as shown in Fig. \ref{Fig_Topo}(a). The lattice constants in the figure are approximately 3.8 \AA\, which is in good agreement with the inter-chalcogen distance. In addition to the lattice, there are a large number of bright spots centred on intermediate sites among the four chalcogens.
From the comparison with the previous neutron diffraction results \cite{WBao,SLi} revealing that the excess Fe atoms reside on the intermediate sites among the four chalcogens, the bright spots in the STM image can be determined to be the excess Fe.

To identify the directions of the $a$- and $b$-axes, we performed a Fourier analysis of Fig. \ref{Fig_Topo}(a), as shown in the inset.
In this analysis, we added to the Fourier transform (FT) image (in the inset) vectors designated $a_{1}$ and $a_{2}$, which correspond to the two directions along the lattice.
In the comparison between the two line cuts of the FT image along the vectors labelled $a_{1}$ and $a_{2}$, we can detect the small but clear difference between the peak positions ($q_{1}$ and $q_{2}$) along the two directions: the peak position ($q_{1}$) along $a_{1}$ is shorter than that ($q_{2}$) along $a_{2}$ by about 2\% (the lattice constant in real space along $a_{1}$ is longer).
From this analysis, we identified the $a_{1}$ and $a_{2}$ directions as the $a$- and $b$-axes, respectively \cite{Piezo}.

\subsection{Typical tunneling spectra}
A tunnelling spectrum averaged over the surface except for the excess Fe sites is shown by the black curve in Fig. \ref{Fig_Topo}(c).
A finite conductance at 0 meV ($E_{\mathrm{F}}$) is consistent with the metallic state in this material below $T_{\mathrm{s}}$.
There is a significant particle-hole asymmetry characterized by the fact that the conductance in negative high energy is greater than that in positive.
In addition to this particle-hole asymmetry, kink structures are observed at $\sim$ +130 meV and $\sim$ $\pm$50 meV as pointed out by the arrows in Fig. \ref{Fig_Topo}(c).
On the excess Fe sites, the conductance at $E_{\mathrm{F}}$ increases and the particle-hole asymmetry becomes more pronounced as shown by the red curve in Fig. \ref{Fig_Topo}(c).
The local enhancement of the conductance at $E_{\mathrm{F}}$ near the excess Fe atoms is consistent with the previous STM experiments in superconducting Fe$_{1+\delta}$Te$_{1-x}$Se$_{x}$ \cite{Hanaguri_2,Kato}.

\begin{figure}[tb]
\begin{center}
\includegraphics[width=7cm]{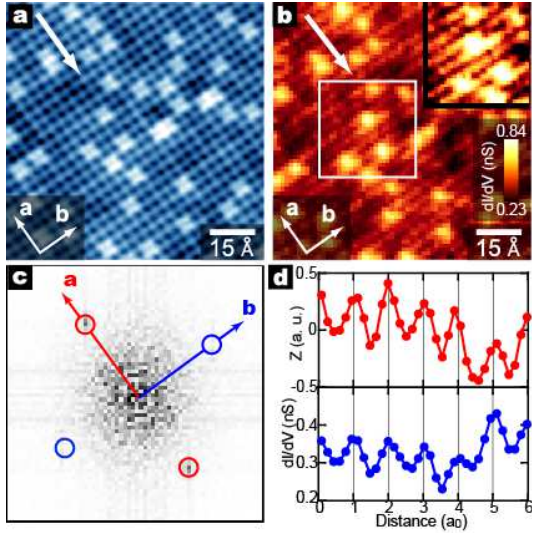}
\end{center}
\caption{
(Color online) (a) An STM image taken at $V_{\mathrm{B}}$ = 500 mV and $I_{\mathrm{set}}$ = 500 pA on an 86-\AA\ square FOV.
(b) Conductance map at 0 meV on the same FOV shown in (a), taken at the set points of $V_{\mathrm{B}}$ = 500 mV and $I_{\mathrm{set}}$ = 500 pA.
Inset of (b) is a high spatial resolution conductance map at 0 meV taken on the region marked by the white box in (a).
This conductance map is drawn by using a higher contrast to display more clearly a unidirectional modulation along $a$-axis in the background of the several bright spots at the excess Fe sites.
(c) FT image of (b).
(d) Line cuts of the topographic signal (red or upper panel) and simultaneously measured conductance (blue or lower panel) along the white arrows in (a) and (b).
}
\label{Fig_SpeGmap}
\end{figure}

\subsection{Unidirectional LDOS modulation}
To investigate the spatial variation of the local density-of-state (LDOS), we performed spectroscopic imaging measurements.
Figure \ref{Fig_SpeGmap} shows an STM image and a conductance map at 0 meV [$G(\boldsymbol{r}, 0\ \mathrm{meV})$ map] with the same FOV as in the STM image.
In the conductance map, there are several bright spots at the excess Fe sites.
In the background of these bright spots, a stripe like or streaky structure running along the $b$-axis is visible as shown in Fig. \ref{Fig_SpeGmap}(b) and more clearly in the inset of Fig. \ref{Fig_SpeGmap}(b): a unidirectional modulation along the $a$-axis exists in this conductance map.
(A clear correlation between the excess Fe atoms and the unidirectional pattern cannot be observed in our measurements).
FT analysis of the $G(\boldsymbol{r}, 0\ \mathrm{meV})$ map also supports the existence of this unidirectional modulation as shown in Fig. \ref{Fig_SpeGmap}(c).
Figure \ref{Fig_SpeGmap}(d) shows the line cuts of the topographic signal (red) and the simultaneously measured conductance (blue) along the white arrow in Fig. \ref{Fig_SpeGmap}(a).
This indicates that the conductance modulation exhibits the same period and phase as the atomic lattice along the $a$-axis.

\begin{figure}[tb]
\begin{center}
\includegraphics[width=6.5cm]{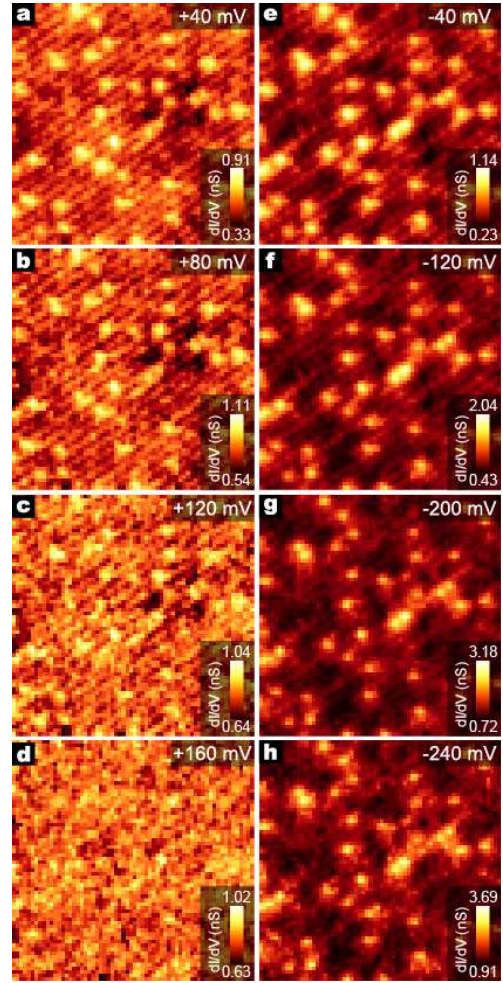}
\end{center}
\caption{
(Color online) Energy dependence of $G(\boldsymbol{r}, E)$ maps. These are taken at $E$ = +40 (a), +80 (b), +120 (c), +160 (d), -40 (e), -120 (f), -200 (g), -240 (h) meV. These maps are taken at the set points of $V_{\mathrm{B}}$ = 500 mV and $I_{\mathrm{set}}$ = 500 pA.
}
\label{Fig_EDepGMap}
\end{figure}

\subsection{Energy evolution of unidirectional LDOS modulation}
\begin{figure}[tb]
\begin{center}
\includegraphics[width=6.5cm]{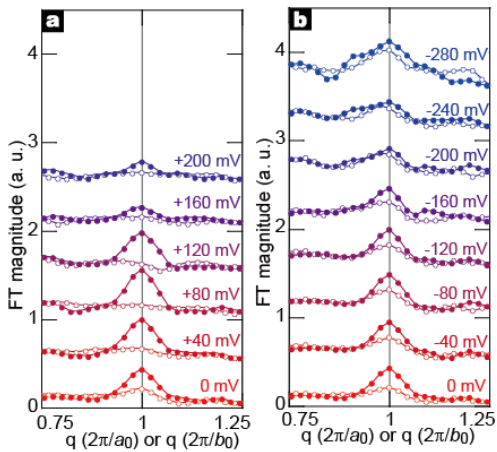}
\end{center}
\caption{
(Color online) Energy dependence of $G(\boldsymbol{q}, E)$ along $a$- (solid circles) and $b$-axes (opened circles) in the positive (a) and negative (b) energy.
Line cuts are shifted vertically for clarity.
The peak position corresponding to the observed unidirectional LDOS modulation does not change with changing the energy.
This energy non-dispersive nature indicates that the observed LDOS modulation is not one of the quasiparticle interference pattern.}
\label{Fig_EDep_FT}
\end{figure}

Figure \ref{Fig_EDepGMap} shows the energy dependence of the conductance maps.
The conductance modulations can be observed from $E$ $\sim$ -200 to +130 meV.
To clarify the energy dependence of the electronic unidirectionality more quantitatively, we plots the FT magnitudes [$G(\boldsymbol{q}, E)$] of the conductance maps along the $a$- (solid circles) and $b$-axes (open circles) as a function of $\boldsymbol{q}$ for different values of energy in Fig. \ref{Fig_EDep_FT}.
The difference between the $G(\boldsymbol{q}, E)$ curves along the $a$- and $b$-axes indicates the unidirectionality of the LDOS modulations.
For positive energy, the difference between the FT magnitudes at $q$ = 2$\pi/a_0$ and 2$\pi/b_0$ gradually increases from $E$ = 0 to +120 meV,  abruptly decreases at +130 meV and is almost undefined above +160 meV.
On the other hand, for negative energy, the difference gradually decreases as the energy decreases from $E$ = 0 to -200 meV, becoming invisible from $E$ = -200 to -280 meV.
Although the physical explanation of this energy asymmetry is not understood at present, this effect will likely be one of the key clues to understand the origin of the unidirectional electronic structure.

\section{Discussion}
\subsection{Relation with underlying orders}
The direction of the observed unidirectional electronic structure coincides with the spin antiparallel direction of the underlying bicollinear AFM order.
Such a unidirectional electronic structure has not been observed in the superconducting Fe$_{1+\delta}$Se$_{x}$Te$_{1-x}$ ($x$ $\sim$ 0.4, 0.15)\cite{Hanaguri_2,Kato} which shows no AFM order.
Therefore, it is presumable that the observed unidirectional structure is closely tied to the underlying AFM order.
However, it remains an open question what kind of mechanisms connects the AFM order (spin state) with the unidirectional electronic structure (electronic state).

The theoretically suggested orbital ordering in the parent phase is one of the important candidates mediating between the spin and electronic states.
In Fe$_{1+\delta}$Te, the structural distortion below $T_{\mathrm{s}}$ lifts the degeneracy of $d_{\mathrm{xz}}$ and $d_{\mathrm{yz}}$ and shifts the energy level of $d_{\mathrm{yz}}$ to be higher than that of $d_{\mathrm{xz}}$.
According to Turner \textit{et al.}, the $d_{\mathrm{xz}}$ orbital is fully occupied by a tightly bound electron whose spin creates the bicollinear AFM order, whereas the $d_{\mathrm{yz}}$ orbital is partially occupied by a mobile electron \cite{Turner}.
Consequently, the movable electron density periodically modulates along the spin anti-parallel direction or the crystal $a$-axis.
Our results indicating the unidirectional electronic structure along the $a$-axis around $E_{\mathrm{F}}$ are consistent with this prediction.

\subsection{Comparison with the other materials}
Here, we compare the unidirectional electronic structures in Fe$_{1+\delta}$Te with those in the 122 and 1111 parent materials\cite{TMChuang,XZhou}.
The direction of the electronic unidirectionality in Fe$_{1+\delta}$Te corresponds to the spin antiparallel direction in the underlying AFM order as seen in the 122 and 1111 systems, although the directions of the AFM order are different from those in the 122 and 1111 systems.
Therefore, the electronic unidirectionality appears to be strongly tied to the underlying orders and to be the inherent phenomenon in the parent phase.
On the contrary, we find several different features in Fe$_{1+\delta}$Te:
(i) No signature clearly indicating that the unidirectional structure is pinned by some impurities could be observed.
(ii) The period in Fe$_{1+\delta}$Te is $a_{0}$ which is considerably shorter than that in CaFe$_{1.94}$Co$_{0.06}$As$_{2}$ (which is approximately 8$a_{0}$).
From these differences, it might be believed that the coupling mechanism of the electronic unidirectionality to the underlying orders in Fe$_{1+\delta}$Te is different from that in CaFe$_{1.94}$Co$_{0.06}$As$_{2}$.

\section{Summary}
In summary, we have investigated the electronic structure lying in the parent state of the iron-chalcogenide superconductor Fe$_{1.07}$Te by using scanning tunnelling spectroscopy.
We find a unidirectional electronic structure with a period of $a_{0}$ along the $a$-axis which coincides with the spin antiparallel direction of the underlying AFM order and the predicted orbital order.
The unidirectional nature is visible within an energy range from about -200 to +130 meV.
Even though the direction of the underlying orders in Fe$_{1+\delta}$Te is different from that in other parent materials, the direction of the unidirectional structure corresponds to the spin antiparallel direction of the underlying AFM order as in Ca(Fe$_{1-x}$Co$_{x}$)$_{2}$As$_{2}$ \cite{TMChuang} and LaOFeAs \cite{XZhou}.
Therefore, it is plausible that the electronic unidirectionality is strongly tied to the underlying orders and is universal and inherent in the nature of the parent state of the iron-based SCs.
The observed electronic unidirectionality provides us an expectation that an in-plane resistivity anisotropy can be found in the detwinned Fe$_{1+\delta}$Te in the same manner for the 122 materials near the parent phase\cite{JHChu}.
On the contrary, we find several differences between the features of the unidirectional structure in Fe$_{1+\delta}$Te and Ca(Fe$_{1-x}$Co$_{x}$)$_{2}$As$_{2}$ \cite{TMChuang}: (i) there is no distinct signature indicating that the unidirectional structure is pinned by some impurities, and (ii) the period is just $a_{0}$ which is different from the value of approximately 8$a_{0}$ in Ca(Fe$_{1-x}$Co$_{x}$)$_{2}$As$_{2}$.
From these differences, the coupling mechanism of the electronic unidirectionality to the underlying orders seems to be different between these materials.
The results in this study provide important hints to understand how the underlying orders couple to the electronic structure in the parent materials of the iron-based superconductors.


\begin{thebibliography}{9}
\bibitem{XtalGrowth}
Mixture of Fe powders and Te grains with nominal molar ratio of Fe$_{1.06}$Te was put into the evacuated quartz tubes and then heated to 1000$^{\circ}$C for 24 hours. After keeping the temperature 1000$^{\circ}$C for 36 hours, the temperature was slowly cooled down to 700$^{\circ}$C at the ratio of -1.5$^{\circ}$C/hour.
\bibitem{Piezo}
Because the difference of the lattice constants between $a$- and $b$-axes is quite small, the calibration of the scanning element (piezo electric tubes) is important to determine the crystal axis.
We performed several STM measurements by using different piezo electric tubes, each of which were calibrated independently.
In each experiments with the different scanning elements, the results shown in this paper were unchanged.
\end{thebibliography}
\end{document}